\documentclass[12pt]{iopart}

\usepackage{graphicx}
\usepackage{iopams}

\begin{document}
\title[Errors in Nuclear DFT]{Error Analysis in Nuclear Density Functional
Theory}

\author{Nicolas Schunck$^1$, Jordan D McDonnell$^1$, Jason Sarich$^2$, Stefan
M Wild$^2$, and Dave Higdon$^3$, }

\address{$^1$ Physics Division, Lawrence Livermore National Laboratory,
Livermore, CA 94551, USA}
\address{$^2$ Mathematics and Computer Science Division, Argonne National
Laboratory, Argonne, IL 60439, USA}
\address{$^3$ Los Alamos National Laboratory, Los Alamos, NM 87545,
USA}

\ead{schunck1@llnl.gov}

\begin{abstract}
Nuclear density functional theory (DFT) is the only microscopic, global approach
to
the structure of atomic nuclei.  It is used in numerous applications, from
determining the limits of stability to gaining a deep understanding of the
formation of elements in the universe or the mechanisms that power stars and
reactors. The predictive power of the theory depends on the amount of physics
embedded in the energy density functional as well as on
efficient ways to determine a small number of free parameters and solve the DFT
equations. In this article, we discuss the various sources of uncertainties and
errors encountered in DFT and possible methods to quantify these uncertainties
in a rigorous manner.
\end{abstract}


\submitto{\JPG}
\maketitle


\section{Introduction}
\label{sec:intro}

The past decade has seen two major developments that have had a profound impact
on nuclear theory. First, progress in nuclear astrophysics has increased the
need for reliable data in neutron-rich and superheavy nuclei. This data has
become a critical input to reaction networks, neutron stars, and supernovae
simulations \cite{kaeppeler2011,langanke2003}. Large uncertainties in
predictions (e.g., of the abundance of elements in the universe) have been
traced to uncertainties in the properties of both ground and excited
states of the nuclei involved during the formation of elements. Second,
targeted programs by funding agencies have fostered
the use of leadership-class computing facilities in science. In particular, the
UNEDF collaboration has deployed high-performance computing (HPC) methods on
various problems in nuclear theory \cite{bogner2013}.

The increased need for reliable, accurate data in systems where no experimental
information is available has shined a spotlight on the predictive power of
nuclear models. It has become increasingly clear that theoretical predictions
must be accompanied with estimates of their error bars. Standard methods of
statistics are mature by now, but often their dissemination has been
hampered by their high computational cost for many problems. In this respect,
the availability of ever more powerful supercomputers and concomitant spread
of HPC techniques can be seen as game changers.

Among nuclear models, density functional theory (DFT) plays a unique role:
it is currently the only global approach to nuclear structure that is
applicable across the whole nuclear chart \cite{bender2003}. It thus features
prominently in high-level applications of nuclear science such as nuclear
fission \cite{younes2011}, superheavy element predictions \cite{hofmann2000},
fundamental symmetries \cite{dobaczewski2005,ban2010}, and neutrinoless
double-beta decay \cite{rodriguez2010}. In all these applications,
high-accuracy,
high-precision DFT calculations are essential, yet relatively few attempts
have been made to estimate the uncertainties underlying these calculations.

In this paper, we discuss some of the issues related to the estimation of
theoretical uncertainties in nuclear DFT. In section \ref{sec:dft}, we discuss
the essential features of the theory. In section \ref{sec:errors},
we analyze various sources of errors of DFT calculations, before
presenting some of the techniques used to propagate these errors in model
predictions in section \ref{sec:uq}. Section \ref{sec:conclusions} gives a
brief summary and outlook.


\section{Nuclear Density Functional Theory}
\label{sec:dft}

In the context of nuclear structure, density functional theory exists in two
variants: the self-consistent mean field (SCMF) theory and the energy density
functional (EDF) theory. We now discuss the main assumptions of both approaches 
and introduce the notation used throughout this paper.


\subsection{Overview of the Theory}
\label{subsec:theory}

The first modern formulation of DFT was from the work of Hohenberg, Kohn, and
Sham on electronic structure theory \cite{hohenberg1964,kohn1965}. Since then, 
considerable effort has been devoted to perfecting the approach, including 
extensions for excited states, statistical ensembles, and time-dependent DFT; 
see \cite{eschrig1996,parr1989} for reviews. The central assumption of DFT is
that a many-body system of electrons interacting via the Coulomb force can be
recast into an independent-particle picture. In this picture, the energy of the
interacting system is expressed as an (unknown) functional of the density of
electrons $n(\bi{r})$. The DFT equations are formally identical to
the Hartree-Fock (HF) equations.

Contrary to many-electron systems, the nucleus is self-bound, and the
interaction is unknown a priori. Until recently the prevailing approach in
nuclear structure has thus been slightly different from electronic DFT. The
starting point has often been an effective nuclear force (or pseudopotential)
$\hat{V}$, which is designed so that, when combined with the Hartree-Fock
approximation to the many-body problem, essential properties of nuclei (e.g.,
saturation density, masses, radii, shell structure) are reproduced. In
essence, this is similar to DFT, except that the energy density $\mathcal{H}$ 
is explicitly derived from the effective potential. This approach is known as 
the self-consistent mean field (SCMF) approach to nuclear structure
\cite{bender2003}.

Scientists have long recognized that pairing correlations are essential
ingredients for explaining the structure of low-lying excited states and 
differences between even and odd nuclei. Given the form of the effective 
interaction $\hat{V}$, the most salient features of pairing are naturally 
incorporated in the SCMF theory by upgrading from the HF to the 
Hartree-Fock-Bogoliubov (HFB) approximation for the ground-state wave function. 
In practice, the energy density now involves an additional term, the pairing 
energy density, which is a functional of the pairing tensor (or pairing 
density) \cite{perlinska2004}.

Guided by the Kohn-Sham and Hohenberg-Kohn theorems (and their generalization
to atomic nuclei \cite{messud2009}), one may therefore
decide to view the energy density $\mathcal{H}$ of the system as the
fundamental unknown of the theory, inasmuch as it is a functional of the
one-body density and two-body pairing tensor $\mathcal{H}[\rho,\kappa]$. This
is the spirit of the energy density functional (EDF) variant of DFT. In
principle, there exists a functional $\mathcal{H}_{0}$ that will give the exact
energy of the system for the solution $\rho, \kappa$ of the HFB equations and
for which {\it all} many-body correlations will be included. Alternatively, one
may decide to retain the effective interaction $\hat{V}$ as the central
component of the theory and improve upon the HF and HFB approximations,
typically by choosing a different ansatz for the nuclear
wave function.


\subsection{Model Parameters and Notations}
\label{subsec:edf}

In this paper, we view both the EDF and SCMF approaches as \emph{models} of
nuclear structure. In practice, they are thus characterized by a finite number,
$n_{x}$, of model parameters $\bi{x} = (x_{1}, \dots, x_{n_{x}})$. In nuclear 
DFT, model parameters are not specified by some underlying theory: they must be 
determined based on some experimental data. By 
$\bi{d} = (d_{1}, \dots, d_{n_{d}})$ we denote the set of $n_{d}$ experimental 
data points used to fit the model. These data points can be of different types 
(e.g., atomic masses, excitation energies, radii, transition strengths); see
section \ref{subsubsec:data}. Typically, $n_{x} \approx 10-20$ for standard 
Skyrme or Gogny EDF, including pairing (see, e.g., the SLy family 
\cite{chabanat1998} or the D1S Gogny force \cite{berger1991}); the number 
of data points varies from about $n_{d}\approx 20$ up to $n_{d} > 2000$ for 
Skyrme or Gogny mass models \cite{goriely2013,goriely2009}.

Given the parameters $\bi{x}$, the value $y_i(\bi{x})$ of an observable 
computed in DFT differs from the experimental value $d_{i}$ by some error 
$\epsilon_{i}$. We collect these calculated observable values and error values 
in $\bi{y} = \bi{y}(\bi{x})$ and $\boldsymbol{\epsilon}$, respectively. In the 
following, we will use Greek indices $\alpha$ to denote parameters, 
$x_{\alpha}$, and Latin indices to denote observables, $d_{i}$ or $y_{i}$.

For the sake of completeness, we mention an important conceptual difference 
between the EDF and SCMF approaches to nuclear structure. In the pure EDF 
approach, the DFT equations are always solved at the HFB level. All information 
about the nucleus is assumed to be encapsulated in the energy density 
$\mathcal{H}$. In principle, it should include both short- and long-range 
correlations, symmetry restoration effects, and so forth. The EDF picture is 
thus characterized by a single model with $n_{x}$ parameters $\bi{x}$. By 
contrast, the model parameters of the SCMF approach are those of the effective 
Hamiltonian. For a given Hamiltonian, however, a hierarchy of approximations is 
available, which reflect the different ansatzes for the nuclear wave function. 
his observation has important  practical consequences when it comes to model 
uncertainties: if the fit of the Hamiltonian parameters $\bi{x}$ is performed 
at a given approximation, say HF, it is not, in principle, applicable at 
another level. Therefore, the fit of an effective force within the SCMF 
approach should always be done at each level of approximation for the 
ground-state wave function. Clearly, this requirement increases considerably 
the difficulty of the task. In the rest of this paper, we will illustrate our 
considerations with studies of the Skyrme functional in the context of the EDF 
rather than the SCMF approach.


\subsection{Sources of Errors}
\label{subsec:errors}

One may distinguish three main sources of uncertainties and errors in nuclear
DFT:
\begin{description}
\item[Model Error -] The most difficult source of uncertainties to
quantify comes from the choice of the energy density. For example, relativistic
EDFs are specified by the number and type of mesons that will be considered and
whether coupling constants will be made density dependent
\cite{niksic2011,ring1996}. Nonrelativistic EDFs may be derived from a
Skyrme-like, zero-range, two-body effective force \cite{vautherin1972} or a
Gogny-like, finite-range, two-body force \cite{decharge1980}, with or without
specific terms such as tensor or generalized density dependencies
\cite{chamel2009,lesinski2007}; alternatively, one may substitute all density
dependencies by effective three-body forces \cite{raimondi2014} or, following
the spirit of electronic DFT, build up the EDF by coupling densities and
currents up to some order \cite{raimondi2014,raimondi2011-a,carlsson2010}.
When EDFs are derived from an effective Hamiltonian, the choice of the ansatz
for the ground-state and excited-state wave functions (HF, HFB, etc.)
introduces additional uncertainties. Quantifying these sources of uncertainties
can be done only on an empirical basis by comparing with experimental data; see
section \ref{subsec:model}. Model errors are examples of systematic errors.
\item[Fitting Bias -] The EDF itself, irrespective of its origin, always
contains a number of free parameters, from about a dozen -- for standard Skyrme 
or Gogny forces -- up to about 30 for the generalized Skyrme forces used in mass
models \cite{chamel2009}. In fact, even for a given EDF, there is sometimes an 
ambiguity about the choice of parameters. For example, the term $\hbar^{2}/2m$, 
in principle a constant, has often be adjusted along the parameters of the 
Skyrme EDF, see Table II in \cite{bender2003}; in relativistic meson-exchange 
EDFs, the mass of some mesons has also been treated as an adjustable parameter 
\cite{ring1996}. The determination of these parameters requires experimental 
data, an optimization algorithm, and a number of specific 
assumptions when solving the DFT equations (spherical or axial symmetry, 
time-reversal symmetry, etc.). Clearly, the choice of the data, the performance 
of the optimization algorithm, and the various hypotheses made in the fit will 
be a source of errors. In the past few years, the nuclear theory community has
expended considerable effort to quantify the resulting uncertainties, which 
can be classified mostly as statistical errors 
\cite{kortelainen2012,kortelainen2010,kortelainen2008}; see section 
\ref{subsec:fit}. 
\item[Numerical Implementation -] Given the form of the EDF and a set of
coupling constants, actual calculations are performed based on a given
numerical implementation. In practice, the DFT equations can be solved in
multiple ways:
\begin{itemize}
\item In a basis formed by, for example, the eigenstates of the harmonic
oscillator (HO) \cite{stoitsov2013,schunck2012}
\item In coordinate space by mesh discretization and direct numerical
integration \cite{bennaceur2005,pei2008}
\item On a lattice \cite{bonche2005}
\item With finite element \cite{poschl1997} or multiwavelet resolution
analysis \cite{fann2009}.
\end{itemize}
Each of these implementations possesses inherent, unavoidable numerical errors. 
Basis expansions are always truncated, possibly inducing a dependence on 
additional parameters (e.g., the oscillator frequency of the HO). The precision 
of mesh or lattice calculations is also contingent on the resolution of the 
underlying grid. 
We give additional examples in section \ref{subsec:numerics}.
\end{description}


\section{Estimating Errors in Nuclear DFT}
\label{sec:errors}

In this section, we provide further details about the various sources of
uncertainties in nuclear DFT calculations. We begin with model errors and
emphasize that they are extremely difficult to completely isolate. We then
present several factors impacting the fit of a given energy density functional.
We close this section with a reminder about numerical errors due to the
particular implementation of the DFT equations.


\subsection{Model Errors}
\label{subsec:model}

As discussed in section \ref{subsec:errors}, model errors in DFT can only be
estimated on an empirical basis by carefully comparing predictions of selected
observables obtained with different functionals. For this comparison to be
meaningful, one should in principle ensure that the optimization
procedure used to determine the coupling constants of the functionals was the
same for all EDFs considered and that the numerical implementation is also
identical. In practice, this is rarely the case.

Mass models are an example where such a comparison is sometimes possible. The
original mass model from the Bruxelles-Montr{\'e}al collaboration was based on
a standard Skyrme force and the BCS approximation \cite{tondeur2000}. In
subsequent iterations of the model, the HFB approximation to the pairing
solution was substituted for the BCS one, more realistic pairing forces were
introduced, and several phenomenological corrections were added. Although the 
experimental data used to perform the fit also evolved over the years, in 
several instances the experimental data, the optimization procedure, and the 
DFT solver used are identical, and only the form of the functional is 
different. For example, in \cite{goriely2013}, the performance of a generalized 
Skyrme force is compared with that of a standard Skyrme force, all things being 
equal.

Another (unpublished) example comes from the UNEDF collaboration and the UNEDF1
protocol \cite{kortelainen2012}. This original parametrization of the Skyrme
EDF was performed at the HFB approximation with an approximate restoration of
particle number using the seniority limit of the Lipkin-Nogami (LN) 
prescription. Details of the fit itself -- choice of data, form of the 
$\chi^{2}$, numerical algorithms, and so on -- can be found in the publications 
by the UNEDF collaboration 
\cite{kortelainen2014,kortelainen2012,kortelainen2010}. The same fit as UNEDF1 
was repeated at the pure HFB level, that is, without the LN prescription. All 
calculations were performed with the HFBTHO solver and the POUNDERS algorithm 
\cite{stoitsov2013,tao-man,SWCHAP14}. Results are summarized in Table 
\ref{table:UNEDF1}.

\begin{table}[!ht]
\caption{Comparison of UNEDF1 parametrization with and without the
Lipkin-Nogami approximate particle number restoration. $\rho_c$ (saturation density) is in
fm$^{-3}$; $E^{NM}/A$ (binding energy per nucleon in nuclear matter), $K^{NM}$ (incompressibility), $a^{NM}_{\mathrm{sym}}$ (symmetry energy coefficient), and
$L^{NM}_{\mathrm{sym}}$ (slope of the symmetry energy) are in MeV; $1/M_s^*$ is dimensionless;
$C_{t}^{\rho\Delta\rho}$ (surface coupling constants) and $C_{t}^{\rho\nabla J}$ (spin-orbit coupling constants), $t=0,1$ are in 
MeV\,fm$^5$; and $V_0^n$ and $ V_0^p $ (pairing strengths) are in MeV\,fm$^3$; for the detail of the 
energy functionals, refer to, e.g., \cite{kortelainen2014}.}
\begin{indented}\item[]
\begin{tabular}{@{}ccc}
\br
Parameters                  & UNEDF1 & UNEDF1-HFB\\
\mr
$\rho_c$                   &    0.158707  &    0.156247  \\
$E^{NM}/A$                 &  -15.80000   &  -15.800000  \\
$K^{NM}$                   &  220.00000   &  244.839379  \\
$a^{NM}_{\mathrm{sym}}$    &   28.986789  &   28.668204  \\
$L^{NM}_{\mathrm{sym}}$    &   40.004790  &   40.109081  \\
$1/M_s^*$                  &    0.992423  &    1.067970  \\
$C^{\rho\Delta \rho}_0$    &  -45.135131  &  -45.599763  \\
$C^{\rho\Delta\rho}_1$     & -145.382168  & -143.935074  \\
$V^n_0$                    & -186.065399  & -234.380010  \\
$V^p_0$                    & -206.579594  & -260.437001  \\
$C^{\rho\nabla J}_0$       &  -74.026333  &  -73.946388  \\
$C^{\rho\nabla J}_1$       &  -35.658261  &  -51.912548  \\
\br
\end{tabular}
\label{table:UNEDF1}
\end{indented}
\end{table}

As anticipated, the major difference between the two parameterizations is in 
the pairing strengths, which are much larger in magnitude when the LN 
prescription is dropped. The two other notable changes are a sizable increase 
of the incompressibility $K^{NM}$ and a sizable decrease of the scalar 
effective mass. The r.m.s.\  deviations for the observables used in the fit are 
listed in Table \ref{table:RMS}. Overall, the original UNEDF1 model, based on 
the HFB+LN approximation, performs better than the UNEDF1-HFB model. Since the 
data used in the fit -- the $\chi^{2}$ objective function, the DFT solver and 
the optimization algorithms -- and even the number of actual parameters, are 
exactly the same in both cases, it is tempting to interpret the differences as 
{\it model differences}. However, this is not exactly true: the optimization is 
always initialized from a given point, and one cannot dismiss a dependence of 
the final result on the starting point, see discussion in \cite{sarich2014} in 
this Focus Issue. In fact, there is no guarantee that the minimum obtained in 
any of these optimizations is the absolute minimum of the 12-dimensional 
hypersurface. Even in this near-ideal calibration, therefore, it is almost 
impossible to completely disentangle {\it model uncertainties} and 
{\it fitting bias}. In the next section, we discuss these fitting biases in 
more detail.

\begin{table}[-ht]
\caption{Root-mean-square deviations for each observable in the UNEDF1
optimization protocol compared for UNEDF1 and UNEDF1-HFB. All r.m.s.\ values
are in MeV, except the ones for proton radii, which are in fm. The acronym OES 
stands for odd-even staggering of nuclear masses.}
\begin{indented}\item[]
\begin{tabular}{@{}lcc}
\br
R.m.s.\            &  UNEDF1  & UNEDF1-HFB  \\ 
\mr
 Deformed masses  &   0.721  &   0.776  \\
 Spherical masses &   1.461  &   1.836  \\
 Proton radii     &   0.016  &   0.022  \\
 OES neutrons     &   0.023  &   0.051  \\
 OES protons      &   0.080  &   0.075  \\
 Fission isomer   &   0.208  &   0.558  \\
\br
\end{tabular}
\label{table:RMS}
\end{indented}
\end{table}


\subsection{Fit of Model Parameters}
\label{subsec:fit}

In practice, parameter estimation comes down to solving an optimization
problem. Given a probability distribution function for the errors
$\boldsymbol{\epsilon}$, a common method of determining model parameter values 
is to maximize the associated likelihood function. For specific choices of the 
distribution, this is equivalent to a least-square fitting, where a $\chi^{2}$ 
function of the squared errors $\boldsymbol{\epsilon}^2$ is minimized.
In this section, we give a brief example of how the form of the $\chi^{2}$ 
function may affect the least-square minimization. 


\subsubsection{Sensitivity on Experimental Data.}
\label{subsubsec:data}

An important feature of DFT is that it is a global approach to nuclear 
structure, which is meant to describe a variety of nuclear properties,
including ground and excited states, and collective motion \cite{bender2003}. 
In order for a functional (or effective interaction) to be truly predictive, 
each parameter thus must be carefully constrained. Doing so could be 
challenging, however, because certain parameters may be especially sensitive to 
specific experimental data and much less sensitive to other data. This fact was 
pointed out in the first systematic studies of uncertainties in DFT by means of 
a singular value decomposition of the model parameters \cite{bertsch2005}. The 
authors found that nuclear binding energies were sensitive only to 3 of the 8 
parameters of a standard Skyrme functional. Similar conclusions were obtained 
when looking at single-particle energies in doubly magic spherical nuclei 
\cite{kortelainen2008}.

In view of these results, the strategy of nuclear mass models, where the single
source of experimental data is atomic masses, may appear too restrictive: while
it will provide excellent agreement with masses, it will do so at the price of
having many parameters ill-constrained. The predictive power of such a model 
for observables that are not masses should be questioned.

In a least-square fit, the usual way to incorporate different data types is to 
take a composite $\chi^{2}$ objective function of the form
\begin{equation}
\chi^{2}(\bi{x}) = \frac{1}{n_{d} - n_{x}} \sum_{t=1}^{n_{T}} \sum_{j=1}^{n_{t}}
\left( \frac{y_{tj}(\bi{x}) - d_{tj}}{\sigma_{t}} \right)^{2},
\label{eq:chi2}
\end{equation}
where $n_{T}$ denotes the number of different data types, $n_{t}$ the number of
data points for type $t$, $n_{d} = \sum_{t} n_{t}$ the total number of data
points over all types, and $n_{x}$ the number of model parameters. The
calculated value of data point number $j$ of type $t$ is denoted by $y_{tj}$, 
with $d_{tj}$ the corresponding experimental value. Because there are
different types of data, relative distances must be properly normalized by the
quantity $\sigma_{t}$ (which, in the general case, could also vary within a 
data type: $\sigma_{tj}$). If we viewed all the data $d_{i} \equiv d_{tj}$ as 
independent random variables following a normal distribution centered about the 
model, $\mathcal{N}(y_{i},\sigma_{i})$, then the quantity defined by
Eq.~(\ref{eq:chi2}) would follow the actual $\chi^{2}$ probability distribution
function. This is the reason that $\sigma_{t}$ is interpreted as an estimate of
the error on the parameter of type $t$. Note that experimental errors on the 
data $\bi{d}$ are not considered here.


\subsubsection{Objective Function.}
\label{subsubsec:chi2}

As shown in Eq.~(\ref{eq:chi2}), the objective function explicitly depends on an
estimate of the choice for the initial errors $\sigma_{t}$ of data type $t$,
and it was shown earlier that this dependence could be significant
\cite{toivanen2008}. In this section, we further illustrate this point in the
realistic setting of the UNEDF1 optimization protocol. As a reference, we take
the UNEDF1-HFB parametrization of Table \ref{table:UNEDF1} and look in
particular at the standard deviation of the odd-even staggering (OES) and of
the fission isomer excitation energy. These two particular data types indeed
appear to be the main drivers of the UNEDF1 parametrization; see fig.~3 in
\cite{kortelainen2012}.

Table \ref{table:RMS_ModWeightsOES} shows the r.m.s.\ deviations of
each observable in the UNEDF1 optimization protocol when the standard deviation
of the OES data varies from 0.025 MeV to 0.100 MeV by a step of 0.025 MeV. For
reference, the UNEDF1-HFB solution, which was obtained with
$\sigma_{OES} = 0.050$ MeV, is also shown. We observe overall large variations,
especially in the reproduction of the masses and the fission isomer excitation
energy. The UNEDF1-HFB obtained with the original weights
probably is very close to giving the best compromise between all different types of observables.

\begin{table}[-ht]
\caption{Root-mean-square deviations for each observable in the UNEDF1
optimization protocol compared for UNEDF1-HFB for a few different values of
the standard deviation $\sigma_{OES}$ (in MeV) of the OES data. All r.m.s.\ 
values are in MeV, except the ones for proton radii, which are in fm.}
\begin{indented}\item[]
\begin{tabular}{@{}ccccc}
\br
                  & $\sigma_{OES}=0.025$ & UNEDF1-HFB &
                  $\sigma_{OES}=0.075$  &  $\sigma_{OES}=0.100$ \\
\mr
 Deformed masses  & 0.960 &  0.776 & 0.764 &  0.775 \\
 Spherical masses & 2.229 &  1.836 & 1.786 &  1.740 \\
 Proton radii     & 0.023 &  0.022 & 0.022 &  0.022 \\
 OES neutrons     & 0.010 &  0.051 & 0.081 &  0.097 \\
 OES protons      & 0.045 &  0.074 & 0.070 &  0.066 \\
 Fission isomer   & 0.830 &  0.558 & 0.488 &  0.441 \\
\br
\end{tabular}
\label{table:RMS_ModWeightsOES}
\end{indented}
\end{table}

The situation is analogous when we vary the standard deviation of the fission
isomer excitation energy, shown in table~\ref{table:RMS_ModWeightsFI}. Here 
again, there are relatively large variations of both the deformed masses and 
the excitation energies. Unfortunately, the two variations are anticorrelated: 
improving the agreement for deformed masses requires an increase in 
$\sigma_{FI}$, which degrades the quality of reproduction of the excitation 
energies. However, the effect is nonlinear: masses are degraded by 36\% when 
dividing $\sigma_{FI}$ by a factor of 2, but improved by a mere 6\% when 
multiplying them by a factor of 2. The sweet spot of the optimization is 
probably somewhere between $0.25$ and $0.50$ MeV.

\begin{table}[-ht]
\caption{Same as Table \ref{table:RMS_ModWeightsOES} for variations of the 
standard deviation $\sigma_{FI}$ (in MeV) of the fission isomer excitation 
energy. }
\begin{indented}\item[]
\begin{tabular}{@{}ccccc}
\br
                  & $\sigma_{FI}=0.25$ &  UNEDF1-HFB  & $\sigma_{FI}=0.75$  &  $\sigma_{FI}=1.00$ \\
\mr
 Deformed masses  & 1.057 &  0.776 & 0.748 &  0.730 \\
 Spherical masses & 1.808 &  1.836 & 1.879 &  1.893 \\
 Proton radii     & 0.023 &  0.022 & 0.021 &  0.021 \\
 OES neutrons     & 0.057 &  0.051 & 0.044 &  0.042 \\
 OES protons      & 0.079 &  0.074 & 0.073 &  0.072 \\
 Fission isomer   & 0.279 &  0.558 & 0.794 &  0.903 \\
\br
\end{tabular}
\label{table:RMS_ModWeightsFI}
\end{indented}
\end{table}

The relatively large variability of optimization results under a change of the
standard deviations $\sigma_{t}$ for each data type is a significant source of
model uncertainties. At this point, there is no magic recipe that would
completely remove them. Bayesian approaches could possibly be used to
provide alternative estimates of these errors.


\subsection{Numerical Implementation}
\label{subsec:numerics}

Of all the possible sources of uncertainties in nuclear DFT computations,
numerical errors stemming from the particular implementation of DFT equations
in a computer code are the easiest to quantify. The vast majority of DFT
solvers are based on the expansion of HF(B) wave functions on a basis. Of
particular interest is the basis made of the eigenstates of the harmonic
oscillator: the HO is a decent approximation of the nuclear mean-field, at
least for deeply bound states; basis functions are analytical; and, most
important, it is the only example where there is an analytical separation
between center of mass and relative motion in a many-body system.

When solving the HFB equations in the HO basis, several approximations can be
imposed on the form of the solutions. In the case of the nonrelativistic
Skyrme pseudopotential, there exist three published, open-source versions of
DFT solvers assuming spherical symmetry \cite{carlsson2010}, axial and
time-reversal symmetry \cite{stoitsov2013}, or no particular symmetry at all 
\cite{schunck2012}. These three solvers have been carefully benchmarked against 
one another and give essentially identical results. A similar package of 
relativistic DFT solvers has recently been published \cite{niksic2014}.

Nuclear DFT calculations in the HO basis are subject to truncation errors.
Because of the finite size of the basis, results become dependent on the
oscillator frequency $\omega_{0}$ or, equivalently, the oscillator length
$b_{0}$. When computing deformed nuclear states, it is convenient to also
deform, or ``stretch,'' the basis states to accelerate convergence: basis
functions are then characterized either by the frequencies
$\bi{\omega} = (\omega_{\perp}, \omega_{z})$ (cylindrical coordinates, 2D) or
$\bi{\omega} = (\omega_{x}, \omega_{y}, \omega_{z})$ (Cartesian coordinates,
3D) or by some spherical-equivalent frequency $\omega_{0}$ and one (or
several) deformation parameters.  See, for example, the discussion in
\cite{stoitsov2013}.

\begin{figure}[ht]
\begin{center}
\includegraphics[width=0.45\linewidth]{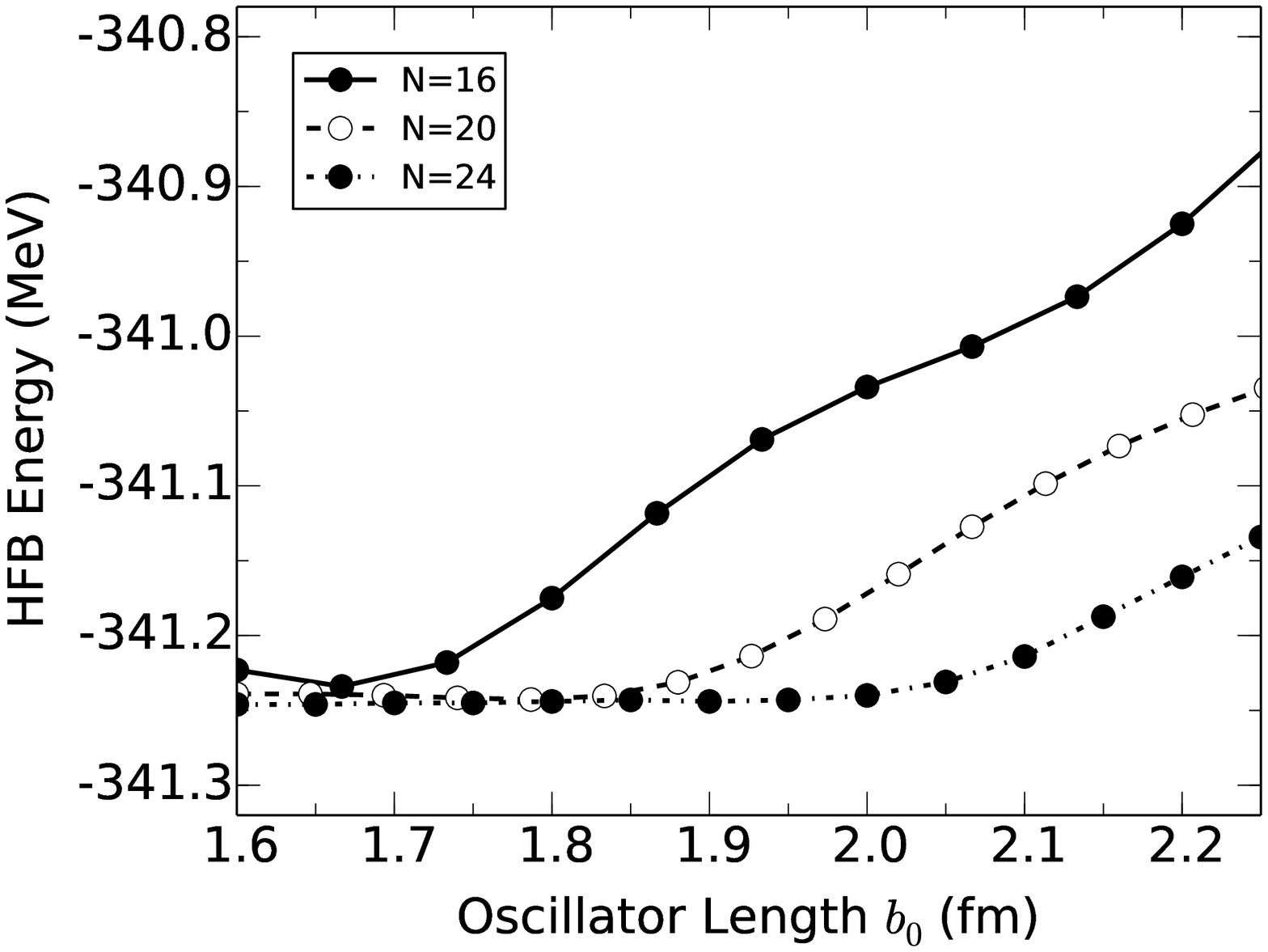}
\includegraphics[width=0.45\linewidth]{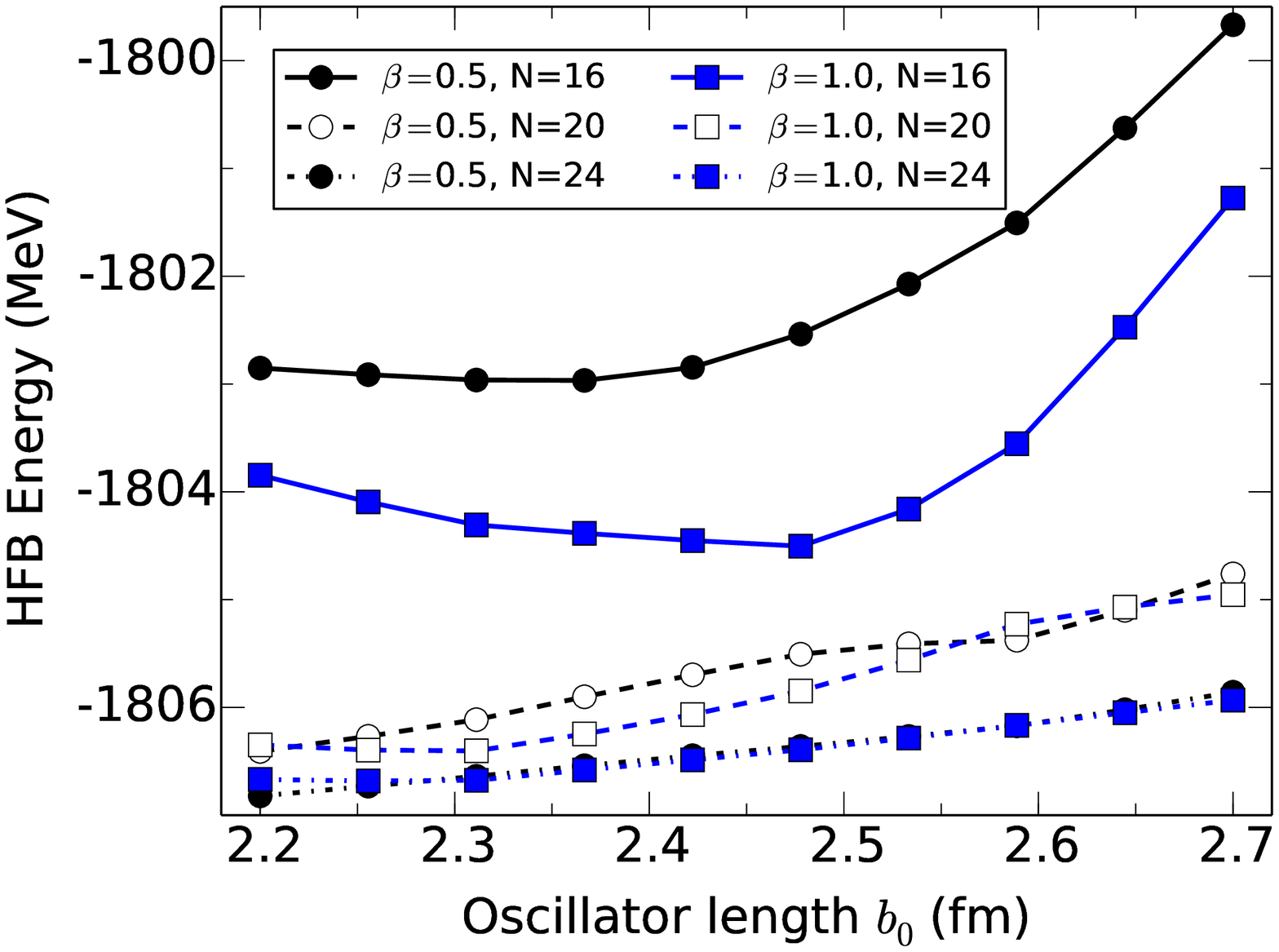}
\caption{Comparison between the pace of convergence of DFT calculations using
HO expansions. Both figures are obtained at the HFB approximation with the SkM*
functional and a surface-volume pairing. Left: Total energy of $^{40}$Ca as
function of the oscillator length $b$ (in fm) for different number of
oscillator shells $N_{\mathrm{shell}}$. Right: Same for the configuration
defined by $\langle \hat{Q}_{20}\rangle = 200$ b and
$\langle \hat{Q}_{40}\rangle = 50$ b$^{2}$ in $^{240}$Pu. Stretched HO bases 
with different deformations $\beta = 0.5$ and $\beta = 1.0$ are used.}
\label{fig:HO}
\end{center}
\end{figure}

This model dependence is illustrated in two extreme cases in fig.~\ref{fig:HO}.
In the left panel, we show the convergence of a simple HF calculation in the 
ground state of $^{40}$Ca; in the right panel, we show the convergence of a 
deformed HFB calculation defined by $\langle \hat{Q}_{20}\rangle = 200$ b and 
$\langle \hat{Q}_{40}\rangle = 50$ b$^{2}$ along the fission path of 
$^{240}$Pu. As expected, the sensitivity on basis parameters is much more 
pronounced in heavier nuclei and for very deformed configurations: across the 
range in values for $b_{0}$, $N_{\mathrm{shell}}$ and the basis deformation 
$\beta$ shown in the figure, the total energy varies by about 0.7 MeV in 
$^{40}$Ca and about 7 MeV in $^{240}$Pu. Including pairing correlations also 
requires larger bases, since higher-lying states may become occupied. In mass 
$A > 200$, one may estimate that ground-state calculations are subject to an 
error greater than 1 MeV even for large HO bases with about $N=20$ full shells; 
this error may reach up to 3-4 MeV in very deformed configurations near the 
scission point \cite{schunck2013}. Early attempts were made to quantify this 
numerical truncation error without performing a full-scale calculation 
\cite{hilaire2007}. Because of the nonlinearity and density dependencies of 
standard EDF, it is not clear whether extrapolation techniques developed in the 
{\it ab initio} community might also be applicable 
\cite{furnstahl2012,coon2012}.

\begin{figure}[ht]
\begin{center}
\includegraphics[width=0.6\linewidth]{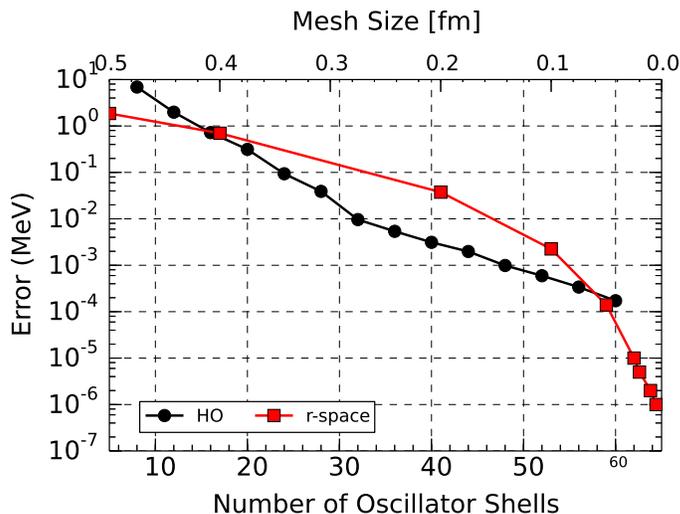}
\caption{Comparison between the pace of convergence of a DFT calculation in
coordinate-space, red squares, and configuration space (HO basis), black 
circles. Results were obtained by setting both direct and exchange terms of the 
Coulomb potentials to 0. The HO basis results are optimized with respect to the
oscillator frequency. Coordinate space calculations were performed with HFBRAD
in a box of 20 fm \cite{bennaceur2005}, HO calculations with HOSPHE
\cite{carlsson2010}.}
\label{fig:convergence}
\end{center}
\end{figure}

DFT calculations performed directly in coordinate space are often considered
more precise than basis calculations. We show in fig.~\ref{fig:convergence} a
comparison between the pace of convergence of a typical HFB calculation in two
different numerical implementations: the direct numerical integration of DFT
equations in coordinate space of the code HFBRAD \cite{bennaceur2005} and the
expansion on spherical HO basis of HOSPHE \cite{carlsson2010}. Calculations
were performed on $^{120}$Sn with the SkM* functional and a simple
surface-volume pairing with a cut-off of $E_{\mathrm{cut}} = 60$ MeV. Since the
Coulomb potential is treated differently in the two codes, it was neglected
here for the sake of comparison. The rate of convergence is approximately
exponential in the HO basis, while it is roughly Gaussian in coordinate space.
Note that in order to achieve the same precision given by a mesh size of
$h = 0.05$ fm (routinely achieved with HFBRAD), nearly 60 {\it full} HO shells
are necessary.

The cost of directly integrating of DFT equations in coordinate space grows
quickly if spherical symmetry is broken. There exist a couple of axial DFT
solvers in coordinate space \cite{pei2008}, but multicore architectures are
essential for reasonable run times. To our knowledge,
there is no full 3D solver in coordinate space. For such arbitrary geometries,
other representations are more promising:
\begin{itemize}
\item The lattice representation of the DFT solvers developed by the
Bruxelles-Bordeaux-CEA collaboration is combined with the imaginary time method
to solve HFB equations \cite{bonche2005}. Its main appeal is that the numerical
precision is essentially independent of the underlying geometry of the nucleus.
\item In a similar spirit, multiresolution wavelet expansions of HFB wave
functions guarantee, by construction, arbitrary precision for observables
\cite{fann2009,pei2012}. Originally developed for quantum chemistry applications, this 
technique has been recently applied to nuclei and has shown great potential for 
complex problems such as fission, highly excited nuclei, or nuclear reactions.
\item A path hitherto neglected in nuclear structure is finite element
analysis. This technique is widely employed in engineering but has only one
application (published) in the context of the relativistic mean field
\cite{poschl1997}.
\end{itemize}


\section{Uncertainty Quantification in Nuclear DFT}
\label{sec:uq}

Most of the discussion presented in section \ref{sec:errors}, especially in
subsections \ref{subsec:model}-\ref{subsec:fit}, was centered on identifying
sources of uncertainties in computations of nuclear properties and sampling
their impact on calculations. For example, comparing predictions of different
energy densities (e.g., Skyrme and Gogny) gives insights into the magnitude
of model uncertainties but does not provide a rigorous metric. The purpose of 
uncertainty quantification in nuclear DFT is to systematically estimate error 
bars in calculations by deploying a variety of statistical and computational 
techniques.


\subsection{Confidence Intervals and Error Propagation}
\label{subsec:error_prop}

The UNEDF collaboration was probably the first to popularize the use of
traditional covariance and sensitivity methods to estimate fitting 
uncertainties \cite{bogner2013}. Starting from a given parameterization
$\bi{x}$ of the model, uncertainties were estimated by using confidence 
intervals \cite{kortelainen2010}. We emphasize that several assumptions are 
made in practice: all the errors $y_{tj}(\bi{x}) - d_{tj}$  are independent of 
one another, each distributed according to a normal distribution (possibly with 
different variances $\sigma^2_t$). Under these conditions, confidence intervals 
can be computed using the central limit theorem result that the estimate 
$\bar{\bi{x}}$ is asymptotically normal, with mean at the true value $\bi{x}$, 
and $n_x \times n_x$ covariance $C(\bi{x})$. We note that the optimization 
process of an EDF yields only the central value $\bar{\bi{x}}$. In order to 
compute the covariance matrix, several additional approximations may be 
invoked.

The most popular of those is that the model outputs $\bi{y(\bi{x})}$ are 
effectively linear with respect to local variations of model parameters 
$\bi{x}$ around the optimized solution $\bar{\bi{x}}$. That is, we can write 
\cite{dobaczewski2014}
\begin{equation}
\bi{y}(\bi{x}) \approx \bi{y}(\bar{\bi{x}}) + \bi{J}(\bar{\bi{x}})(\bi{x} - \bar{\bi{x}})
\end{equation}
where $\bi{J}$ is the sensitivity matrix at $\bar{\bi{x}}$,
$
J_{\alpha i} (\bar{\bi{x}})
=
\left(\begin{array}{c} \displaystyle
\frac{\partial y_{i}(\bar{\bi{x}})}{\partial x_{\alpha}}
\end{array}\right).
$
Under this assumption, one can show that the covariance matrix reduces to
\begin{equation}
C = \sigma^2 \left( \bi{J}(\bar{\bi{x}}) \bi{J}(\bar{\bi{x}})^{T} \right)^{-1} .
\end{equation}
Note that the covariance matrix thus defined reflects both fitting errors, 
since it depends on the optimal parameters $\bar{\bi{x}}$, and model errors 
through the use of the variance $\sigma$. It is another illustration of the 
difficulty to completely disentangle these two sources of errors. 
Alternatively, one can use the Hessian of the $\chi^2$ to estimate the
covariance matrix for $\bar{\bi{x}}$
\[
C = H^{-1},\,  \mbox{ where }
H_{\alpha\beta}(\bar{\bi{x}}) =
\left(\begin{array}{c} \displaystyle
\frac{n_{d} - n_{x}}{2}\frac{\partial^{2}\chi^{2}}{\partial x_{\alpha}\partial
x_{\beta}} (\bar{\bi{x}})
\end{array}\right).
\]
Uncertainties on $\bar{\bi{x}}$ induce uncertainty on the model prediction so 
that the variance of a computed observable $y_{j}(\bar{\bi{x}})$ is simply 
given by
\begin{equation}
 \mbox{Var}(y_j(\bar{\bi{x}})) = \sum_{\alpha\beta} J_{\alpha j} C_{\alpha\beta}
J_{\beta j}.
\end{equation}
This covariance technique has been recently applied to estimate the
information content of the electric dipole strength \cite{reinhard2013}, the
correlation between electric dipole polarizability and neutron skin
\cite{piekarewicz2012}, the uncertainties on the weak charge form factor
\cite{reinhard2013-a}, and the neutron skin of neutron-rich nuclei
\cite{kortelainen2013}.

\begin{figure}[ht]
\begin{center}
\includegraphics[width=0.45\linewidth]{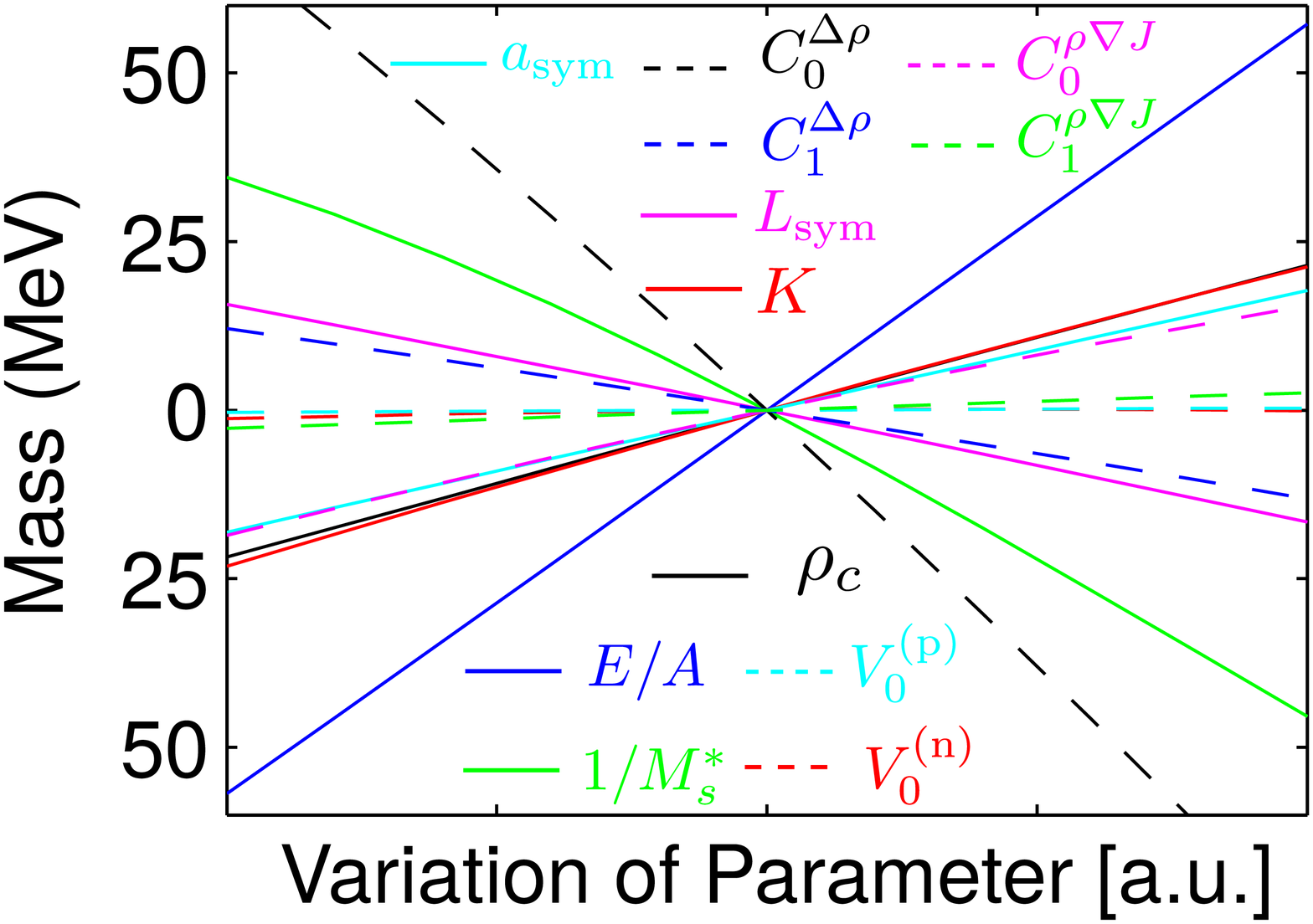}
\includegraphics[width=0.45\linewidth]{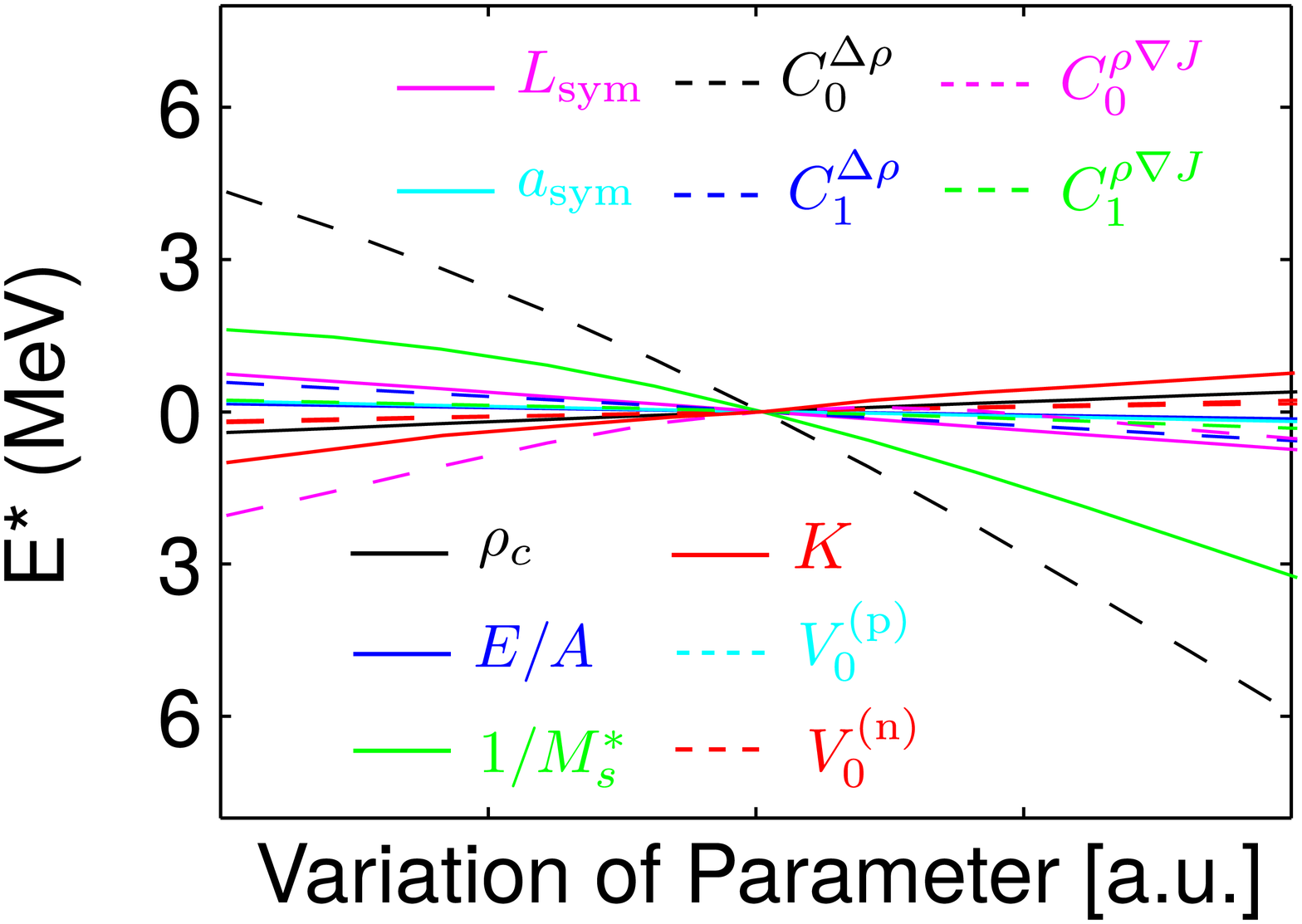}
\caption{Variation of the mass (left panel) and the excitation energy of the
fission isomer (right panel) of $^{240}$Pu as a function of each of the 12
parameters of the UNEDF1 functional. All parameters have been normalized to the 
interval of variation listed in Table II of \cite{kortelainen2012}, column 
marked ``Bounds.'' }
\label{fig:linearity}
\end{center}
\end{figure}

The two approximations of independent errors and of linearity near the solution
seem the strongest. In fig.~\ref{fig:linearity}, we show the variations of the 
mass of the $^{240}$Pu nucleus and of the excitation energy of its fission 
isomer as a function of each of the twelve parameters of the UNEDF1 functional. 
Each parameter has been normalized between 0  and 1 based on the ``reasonable 
interval'' of variation listed in  \cite{kortelainen2012}. We see that the mass 
behaves nearly linearly across the entire parameter range; by contrast, the 
excitation energy of the fission isomer shows some marked deviations, in 
particular as a function of the scalar effective mass $M_{s}^{*}$, the 
isoscalar surface, and the spin-orbit terms $C_{0}^{\rho\Delta\rho}$ and 
$C_{0}^{\rho\nabla J}$. Since these deviations are more pronounced as we go 
away from the local solution given by UNEDF1, they may not impact the 
calculation of the covariance matrix and the related estimate of uncertainties; 
there are other instances of data used in EDF fits, such as single-particle 
energies in doubly-magic nuclei, were non-linearities are large enough to be 
noticeable \cite{toivanen2008,kortelainen2008}. 


\subsection{Application of Bayesian Statistics in Nuclear DFT}
\label{subsec:bayes}

The Bayesian approach is based on Bayes' theorem for conditional probabilities:
the probability that a continuous alternative $A$ lies between $A$ and
$A+dA$ given $B$ and $C$ is given by
\begin{equation}
p(A|BC)dA = \frac{p(B|AC)p(A|C)dA}{\int p(B|AC)p(A|C)dA}.
\end{equation}
Based on this theorem, the Bayesian approach consists of treating the model
parameters $A$ as genuine random variables depending on some data $B$ and some
other circumstances $C$. The goal is to find the probability distribution
function (p.d.f.) of the random variable $A$, that is, the probability of 
having $A$ between $A$ and $A+dA$ given $B$ and $C$. From the Bayesian 
perspective,  uncertainty regarding the fixed, but unknown, model parameters is 
described  with probability. In practice, Bayes' theorem is also often given as
\begin{eqnarray*}
\mathrm{posterior} &\propto& \mathrm{likelihood} \times \mathrm{prior}, \qquad
\mbox{or} \\
\pi(\bi{x}|d) &\propto& \mathcal{L}(\bi{x},d) \times p(\bi{x}).
\end{eqnarray*}
In the context of nuclear DFT, this equation should be interpreted as follows. 
We start with a prior density $p(\bi{x})$ for the model parameters $\bi{x}$. By
default, one may assume a uniform distribution between some sensible,
physically-motivated intervals so that $p(\bi{x}) \propto I[\bi{x} \in C]$,
where $C$ denotes the $n_{x}$-dimensional prior rectangle, $n_{x} = 12$ for the 
UNEDF1 protocol. Based on a set of experimental data $\bi{d}$, we then define 
our likelihood function $\mathcal{L}(\bi{x},d)$ based on some 
$\chi^{2}(\bi{x},\bi{d})$ function, since, for normally distributed random 
variables, we know that 
$\mathcal{L}(\bi{x},\bi{d}) \propto p(\bi{d}|\bi{x}) 
\propto e^{-\chi^{2}(\bi{x},\bi{d})}$. 
The posterior distribution will be estimated by sampling from the posterior 
density of the parameters, $\pi(\bi{x}|d)$.

The Bayesian approach offers numerous advantages. It provides a full
probabilistic description of the model parameters, allowing very general
dependence between model parameters, from which one may deduce the mean, 
standard deviation, and covariance matrix, if desired. It can easily
incorporate the impact of new data: a posterior distribution obtained from a
given set of data can serve as a prior distribution if the dataset is extended
(or modified). On the other hand, the computational cost of building a full
p.d.f.\ in the context of nuclear DFT can be significant. In the case of the
UNEDF parameterizations, each $\chi^{2}$ function involves on the order of 100
deformed HFB calculations, each taking on the order of 5-10 minutes. The
parameter space has dimension 12: in such a space, Markov chain Monte Carlo
techniques, which are often used to build the posterior distribution, could
easily require dozens of thousands of iterations before convergence. Various
techniques can be deployed to mitigate this cost, such as the construction of
metamodels (or response functions, or emulators) for the $\chi^{2}$ function. 
This topic is discussed in greater details in \cite{higdon2014} in this Focus 
Issue.


\section{Conclusions}
\label{sec:conclusions}

In this paper, we have discussed the various sources of errors and 
uncertainties in nuclear density functional theory. In particular, we have 
distinguished between model errors, fitting errors, and numerical 
implementation errors. Implementation errors are purely statistical and, in
principle, are the easiest to control, although they can become significant in 
specific applications such as fission, where systems become extremely 
elongated, or neutron-rich nuclei near or beyond the drip lines, where the 
coupling to continuum becomes significant. Considerable work was recently 
devoted to estimating and propagating fitting errors, mostly through covariance 
techniques. Such errors are mostly of statistical nature -- in the sense that 
more data would improve the fit. However, we also argued that it is difficult 
to isolate fitting errors from the intrinsic uncertainties of the model. 
Indeed, model errors are unavoidable in the theoretical description of any 
quantum many-body problem, in particular in the nuclear physics case where the 
interaction is not known. These systematic errors are by far the most difficult 
uncertainties to estimate, and preliminary indications are that they can be 
very large. We have also emphasized that Bayesian techniques may represent a 
promising path toward a more complete quantification of uncertainties in DFT. 
These methods are computationally costly but can be deployed in a variety of 
settings.


\section*{Acknowledgment}
This work was partly performed under the auspices of the U.S.\ Department of
Energy by Lawrence Livermore National Laboratory under Contract
DE-AC52-07NA27344. It was supported by the SciDAC activity within the U.S.\ 
Department of Energy, Office of Science, Advanced Scientific Computing Research
under contract number DE-AC02-06CH11357. Computational resources were
provided through an INCITE award ``Computational Nuclear Structure'' by the
National Center for Computational Sciences (NCCS) and National Institute for
Computational Sciences (NICS) at Oak Ridge National Laboratory, through an
award by the Livermore Computing Resource Center at Lawrence Livermore National
Laboratory, and through an award by the Laboratory Computing Resource Center at
Argonne National Laboratory.


\section*{References}

\bibliographystyle{unsrt}
\bibliography{../bibs/jphg,../bibs/books}

\end{document}